\documentclass[twocolumn,showpacs,preprintnumbers]{revtex4}
\usepackage{amssymb}
\usepackage{amsfonts}
\usepackage{amsmath}
\usepackage{graphicx}
\usepackage{dcolumn}
\usepackage{bm}

\input{tcilatex}

\begin{document}

\title{Simple Quantum Model of Learning Explains the Yerkes-Dodson Law in
Psychology}
\author{E. D. Vol}
\email{vol@ilt.kharkov.ua}
\affiliation{B. Verkin Institute for Low Temperature Physics and Engineering of the
National Academy of Sciences of Ukraine 47, Lenin Ave., Kharkov 61103,
Ukraine.}
\date{\today }

\begin{abstract}
We propose the simple model of learning based on which we derive and explain
the Yerkes-Dodson law - one of the oldest laws of experimental psychology.
The approach uses some ideas of quantum theory of open systems (QTOS) and
develops the method of statistical description of psychological systems that
was proposed by author earlier.
\end{abstract}

\maketitle

\section{Introduction}

The Yerkes-Dodson Law (YDL) is one of the oldest laws of psychology of
motivation. It was discovered experimentally more than one hundred years ago
in 1908 \cite{1a}.\ The main goal of original Yerkes-Dodson experiments was
to examine to what extent motivation and emotion arousal have an influence
on learning process \ and facilitate acquisition of habits which are needed
for individual \ in solving everyday problems. The main result\ \ \ obtained
by Yerkes and Dodson as well as similar more late experiments with various
species such as rats, chickens, cats and also with \ men (with specific
tasks for each species of course) turned out to be universal. In all cases
there is optimal level of arousal which provides the most successful
execution of required task. Besides the hallmark of the YDL is the fact that
optimal level of arousal depends on the complexity of the task, namely, more
complex the task the lower optimal level of arousal should be.\ In classical
handbooks of experimental psychology, (see e.g. \cite{2b}\ )\ as the verbal
explanation of the YDL usually the following argument was drawn: since
excessive raising of arousal leads to essential disorganization of behavior
of individual therefore the execution of task in such situation should be
worse.

Althouh such argument seems to be quite reasonable but a physicist who is
interested in psychology undoubtedly would like to have some explicit model
of learning that\ taking into account influence of arousal on the process
allows him to\ derive (from clearly articulated assumptions ) the YDL
together with its hallmark. This is just the main goal of present paper. The
rest of the paper is organized as follows. In Sect.2 \ we briefly remind the
basic points of statistical method for the description of simple
psychological phenomena which was proposed by author earlier \cite{3b}. This
method based on some ideas of QTOS is the starting tool for our further
analysis. In Sect.3 we formulate simple phenomenological model of learning
and discuss to what extent it can be considered as reasonable from
psychological point of view. In Sect.4 we demonstrate that the model
proposed let one rigorously derive the YDL together with its hallmark. In
conclusion we are discussing the possibility of experimental verification of
the model proposed. Now let us go to the detail presentation of the paper.

\section{Preliminary Information}

In this part we want to remind the main points of the method which was
proposed by author earlier \cite{3b} for the statistical description of
simple phychological phenomena (both individual behavior and group
processes). First \ of all we want to emphasize that the question is only
about statistical description. The fact is that due to extraordinary
complexity of various psychological phenomena according to views of many
modern psychologists see e.g. \cite{4b} ) just the statistical approach to
behavior description is the most relevant. Therefore we will assume the task
of behavior description is solved if we can find the probabilities of all
psychological states which are possible in situation under study. We believe
that for this purpose analogy with QTOS where such aprroach is inherent from
the very beginning may be very productive (for more detail argumentation in
favour of this analogy see \cite{3b}). As one knows in QTOS fundamental
concept of density matrix of the system $\rho $ let one correctly describe
such state of affairs. Later in this paper we use the analogy between
psychology and QTOS only at phenomenological level and we will show that
this analogy significantly helps one in explaining the YDL. Note the only
information from QTOS that we needed for our purpose is the Lindblad
equation. This equation derived in QTOS under rather general restrictions
imposed on the system of interest (see \cite{5b}) let one explicitly
describe the required density matrix evolution in time, that is the
"behavior" of the system.

The Lindblad equation in QTOS has the following form \cite{5b}:%
\begin{equation}
\frac{\partial \widehat{\rho }}{\partial t}=-\frac{i}{\hbar }\left[ \widehat{%
H},\widehat{\rho }\right] +\sum\limits_{i}\left[ \widehat{R}_{i}\widehat{%
\rho },\widehat{R_{i}^{+}}\right] +\ \ h.c.  \label{k1}
\end{equation}

In Eq. (\ref{k1}) $\ \widehat{H}$ \ is hermitian operator ("hamiltonian" of
open system) and $\widehat{R_{i}}$ are a set of nonhermitian operators that
specify all links of open system in question with its environment. If
initial state of the system $\rho (0)$ is known Eq. (\ref{k1}) let one to
find all desired information about the system at any time. In order to apply
Eq. (\ref{k1}) to concrete psychological system of interest we must firstly
define its possible psychological states and in addition to define the form
of all operators $\widehat{R_{i}}$ incoming in Eq. (\ref{k1}). To this end
we must know all stimuli i.e. all psychological forces acting on the system.
In other words applied to our problem to explain the YDL within the
framework of Eq. (\ref{k1}) we must formulate relevant mathematical model of
the learning process taking into account the influence of arousal on this
process. Let us now begin to adress this issue.

\section{Phenomenological model of learning}

In present paper we offer as a basis the next phenomenological model of
learning. We assume that the space of psychological states of trained
individual (its life space according terminology of K. Lewin's psychological
field theory \cite{6b}) can be represented as linear superposition of three
basic states: state $\left\vert 1\right\rangle \equiv \left( 
\begin{array}{ccc}
0 & 0 & 1%
\end{array}%
\right) ^{T}$ - that corresponds to untrained individual, state $\left\vert
2\right\rangle \equiv \left( 
\begin{array}{ccc}
0 & 1 & 0%
\end{array}%
\right) ^{T}$- pre-trained \ state (in this state individual has acquired
some elements and features of the habit , but the skill for the task
completition still lacks) and the trained state $\left\vert 3\right\rangle $ 
$\equiv \left( 
\begin{array}{ccc}
1 & 0 & 0%
\end{array}%
\right) ^{T}$ in which individual is able successfully execute required
task. Proposing such model we have in mind as correct the hypothesis of
functional equvalence between perception process and other cognitive
processes. Remind that according to modern cognitive psychology (see e.g. 
\cite{7b}) sensory information processing in brain especially visual
information processing consists of two consecutive stages: the first stage
(segmentation) is the formation of coherent clusters of perception at which
groups of similar features of the perceived object combine together and the
second stage (binding) on which fragments of perception are integrated into
complete image. In fact our model of learning assumes that state $\left\vert
2\right\rangle $ corrresponds to segmentation stage and state $\left\vert
3\right\rangle $ to the binding stage. In addition we believe the learning
process can be represented as the set of four transitions between basic
states: a) transition from $\left\vert 1\right\rangle $ to $\left\vert
2\right\rangle $- process of pre-learning b) transition from $\left\vert
2\right\rangle $ to $\left\vert 3\right\rangle $- process of habit
acquisition , c) transition from $\ \left\vert 3\right\rangle $ to $%
\left\vert 2\right\rangle $-process of partial loss of the habit due to the
influence of different noise and d) transition from $\left\vert
3\right\rangle $ to $\left\vert 1\right\rangle $- natural process of
"forgetting" of the habit when it is not claimed for a long time. We believe
that all these transitions can be described on the basis of the Lindblad
equation with the help of four relevant operators $\widehat{R_{i}}$.
Speaking more precisely: transition a) can be described by the operator $%
\widehat{R_{1}}=\sqrt{\frac{a}{2}}%
\begin{pmatrix}
0 & 0 & 0 \\ 
0 & 0 & 1 \\ 
0 & 0 & 0%
\end{pmatrix}%
$, transition b)- by the operator $\widehat{R_{2}}=\sqrt{\frac{b}{2}}%
\begin{pmatrix}
0 & 1 & 0 \\ 
0 & 0 & 0 \\ 
0 & 0 & 0%
\end{pmatrix}%
$ , transition c) by the operator $\widehat{R_{3}}=\sqrt{\frac{c}{2}}%
\begin{pmatrix}
0 & 0 & 0 \\ 
1 & 0 & 0 \\ 
0 & 0 & 0%
\end{pmatrix}%
$ and transition d) by the operator $\widehat{R_{4}}=\sqrt{\frac{d}{2}}%
\begin{pmatrix}
0 & 0 & 0 \\ 
0 & 0 & 0 \\ 
1 & 0 & 0%
\end{pmatrix}%
$. Coefficients \ $a$,$b$,$c$, $d$ incoming in operators $\widehat{R_{i}}$
reflect the power of corresponding transitions.Probabilities $\rho _{1\text{ 
}}$, $\rho _{2}$, $\rho _{3}$ of finding the individual in one of basic
states in learning process one can consider as diagonal elements of density
matrix $\widehat{\rho }=%
\begin{pmatrix}
\rho _{1} & 0 & 0 \\ 
0 & \rho _{2} & 0 \\ 
0 & 0 & \rho _{3}%
\end{pmatrix}%
$.

In the paper \cite{3b} it \ was shown that if at initial time density matrix 
$\widehat{\rho \left( 0\right) }$ has diagonal form then under certain
restrictions on the form of operators $\widehat{R_{i}}$ such diagonal form
will be conserved in time. It is easy to check that with above- mentioned
choice of $\ $operators $\widehat{R_{i}}$ in the model just this case
occurs. Now having in hands all relevant tools we can move to a direct
derivation of the YDL .

\section{Derivation of the Yerkes -Dodson Law}

In this part we assume that the main features of learning process efficiency
in the light of arousal influence can be correctly described in the
framework of foregoing model with the help of the Lindblad equation Eq. (\ref%
{k1}). Substituting in it operators $\widehat{\rho }$ and $\widehat{R_{i}}$
\ $(i=1,2,3,4)$ in the form specified above after simple algebra we obtain
the next system equations of motion for required probabilities $\rho _{1}$, $%
\rho _{2}$,$\rho _{3}$:%
\begin{eqnarray}
\frac{d\rho _{1}}{dt} &=&-a\rho _{1}+d\rho _{3}  \notag \\
\frac{d\rho _{2}}{dt} &=&a\rho _{1}-b\rho _{2}+c\rho _{3}  \label{k22} \\
\frac{d\rho _{3}}{dt} &=&b\rho _{2\text{ \ \ }}-\left( c+d\right) \rho _{3} 
\notag
\end{eqnarray}

In the remainder of the paper we are interested only in stationary states of
individual who trains therefore in Eq. (\ref{k22}) we equate all $\frac{%
\partial \rho _{i}}{\partial t}$ to zero and taking into account
normalization condition $\rho _{1}+\rho _{2}+\rho _{3}=1$ obtain expressions
for required probabilities $\rho _{i}$:%
\begin{eqnarray}
\rho _{1} &=&\frac{bd}{a\left( b+c+d\right) +bd};  \notag \\
\rho _{2} &=&\frac{a\left( c+d\right) }{a\left( b+c+d\right) +bd};
\label{k31} \\
\rho _{3} &=&\frac{ab}{a\left( b+c+d\right) +bd}.  \notag
\end{eqnarray}

Note that until now we did not take into account in explicit form influence
of arousal \ on learning process. At this point we are going to do it.
Because of psychological reasons it can be assumed that increase of arousal
facilitates the transitions a) and c) and has negligible effect on
transitions b) and d). Therefore the easiest way to take into account
arousal level in our model is to put $c=at$ and then to consider
coefficients $a$,$b,d,t$ as independent parameters. Now we believe that
value of $a$ exactly determines the arousal level of individual in the
learning process. Let us see that there is optimal level of arousal for
successful training in the model. Indeed if we write the extremal condition
for probability of trained state $\rho _{3}$ that reads as $\frac{\partial
\rho _{3}}{\partial a}=0$ we obtain that $a_{ext}^{2}=\frac{bt}{d}$. Clearly
this value of $a$ corresponds exactly to the optimal level of arousal which
facilitate the task realization. Besides the hallmark of the YDL is
immediately implies from above expression for $a_{ext}$ \ because it is
clear more difficult task than the less coefficient $b$ should be.

In conclusion let us discuss briefly the possibility to verify
experimentally the validity of learning model proposed and hence
(indirectly) the reasonableness of original analogy between psychology and
QTOS. To this end one must reproduce experiments similar to Yerkes and
Dodson but using statistically significant ensemble of learning individuals.
In this situation it is possible to get reliable values of probabilities $%
\rho _{1}$,$\rho _{2}$,$\rho _{3}$and to check the model proposed. In fact
there is some problem of checking expressions Eq. (\ref{k31}) as long as
values of phenomenological coefficients $a,b,c,d$ are unknown. To avoid this
obstacle it is useful to find some relations that would not depend on values
of these coefficients. To this end let us write down expressions Eq. (\ref%
{k31})in the optimal point: $a^{2}=\frac{bd}{t}$ where they take the next
form

\begin{eqnarray}
\rho _{1} &=&\frac{\sqrt{bdt}}{b+d+2\sqrt{bdt}};  \notag \\
\rho _{2} &=&\frac{d+\sqrt{bdt}}{b+d+2\sqrt{bdt}};  \label{k4} \\
\rho _{3} &=&\frac{b}{b+d+2\sqrt{bdt}}.  \notag
\end{eqnarray}

We see that in optimal point the unequality: $\rho _{2}\geq \rho _{1\text{ \ 
}}$ always holds for any values of $b,d,t$. If we assume in addition that
ratio $\frac{d}{b}\ll 1$( what means that forgetting of acquired habit is
rather slow) than we obtain from Eq. (\ref{k4}) approximate equality:$2\rho
_{1}+\rho _{3}\approx 1$ or $\rho _{1}\approx \rho _{2}$. Thus we come to
the conclusion: if $\rho _{1},\rho _{2},\rho _{3}$ \ can be measured
directly in statistical learning experiment we have actual possibility to
verify the validity of model proposed.

Let us summing up the results of our consideration. Starting from attractive
analogy between QTOS and based on simple statistical model of learning we
obtain expressions for chances of individual to reach one or another level
of task execution \ under the conditions of changing arousal. We have shown
that this approach let one strictly derive the Yerkes-Dodson Law as well as
its hallmark. We would like to express a hope that new experiments help to
confirm (or to disprove) the statistical approach to learning process
proposed in present paper.

\end{document}